\documentclass{PoS}
\title{Recent results of the DANSS experiment}

\ShortTitle{Recent results of the DANSS experiment}

\author{\speaker{M.~Danilov }\\
        Lebedev Physical Institute of the Russian Academy of Sciences,\\ 
        53 Leninskiy Prospekt, Moscow, 119991, Russia\\ 
        E-mail: \email{danilov@lebedev.ru}}

\author{On behalf of the DANSS Collaboration\\}
\def\anti_particle{\tilde}
\def\cls{CL$_s$ }

\abstract{ We present new results of the DANSS experiment on the searches for sterile neutrinos. They are based on 2.1 million of inverse beta decay events collected at 10.7, 11.7 and 12.7 meters from the reactor core of the 3.1 GW Kalinin Nuclear Power Plant in Russia. This data sample is 2.5 times larger than the data sample in the previous DANSS publication. The search for the sterile neutrinos is performed using the ratio of $\anti_particle\nu_e$ spectra at two distances. This method is very robust against systematic uncertainties in the $\anti_particle\nu_e$ spectrum and the detector efficiency. We do not see any statistically significant sign for the $\anti_particle\nu_e$ oscillations. This allows us to exclude further a large and interesting part of the sterile neutrino parameter space. 
 All results are preliminary.\\}

\FullConference{
European Physical Society Conference on High Energy Physics - EPS-HEP2019 -\\
			10-17 July, 2019\\
			Ghent, Belgium}

\begin{document}
\section{Introduction}

The number of active neutrinos is limited to 3 by the measurements of the Z boson decay width \cite{PDG}.
 However, the existence of additional sterile neutrinos is not excluded. There are even several indications of their existence. 
The deficit of neutrino events in the calibration runs of the GALEX and SAGE Gallium experiments \cite{SAGE} (Galium anomaly (GA)) , as well as the deficit of the $\anti_particle\nu_e$ flux from reactors \cite{Mueller} (Reactor antineutrino anomaly (RAA)), can be explained by active-sterile neutrino oscillations at very short distances requiring a mass-squared difference of the order of 1~eV$^2$ \cite{Mention2011}.
Recently the MiniBooNE collaboration observed electron (anti)neutrino appearance in the muon (anti)neutrino beams. The significance of the effect reaches 6.0$\sigma$ level when combined with the LSND result \cite{MiniBooNE}. Even more recently the NEUTRINO-4 collaboration claimed the observation of electron $\anti_particle\nu_e$ oscillations to sterile neutrinos with a significance of about 3$\sigma$ \cite{Neutrino-4}. If these results are confirmed, New Physics beyond the Standard Model would be required.
On the other hand, the DANSS experiment \cite{DANSS-PLB} and several other reactor experiments at short baselines obtained quite strict limits on the hypothetical sterile neutrino parameters. For a recent review see e.g. \cite{DanilovICPPA}. We present here new results of the DANSS experiment on the searches for sterile neutrinos based on 2.1 million $\anti_particle\nu_e$ events. This data sample is 2.5 times larger than in the previous publication~\cite{DANSS-PLB}.

The survival probability of a reactor $\anti_particle\nu_e$ at short distances in the 4$\nu$ mixing scenario (3 active and 1 sterile neutrino) is described by a familiar expression

\begin{equation}
1-\sin^22\theta_{14}\sin^2\left(\frac {1.27\Delta m_{14}^2 [\mathrm{eV}^2] L[\mathrm m]}{E_\nu [\mathrm{MeV}]}\right).
\end{equation}

The existence of sterile neutrinos would manifest itself in distortions of the $\anti_particle \nu_e$ energy spectrum at short distances.  The most reliable way to observe such distortions is to measure the $\anti_particle \nu_e$ spectrum with the same detector at different distances. In this case, the shape and normalization  of the $\anti_particle \nu_e$ spectrum as well as the detector efficiency drop out. Detector positions should be changed frequently enough to cancel out time variations of the detector and reactor parameters. The DANSS experiment uses this strategy and measures $\anti_particle \nu_e$ spectra at 10.7~m, 11.7~m, and 12.7~m from the reactor core center to the detector center. 
The detector positions are changed usually 3 times a week. Antineutrinos are detected by means of the Inverse Beta Decay (IBD) reaction
\begin{equation}
\label{eq1}
\anti_particle{\nu}_e + p \rightarrow e^+ + n ~\mbox{with}~ E_{\anti_particle\nu} = E_{e^+} + 1.80~\mathrm{MeV}.
\end{equation}

\section{The DANSS detector and data analysis}
The DANSS detector is described elsewhere \cite{DANSS}. Here we mention only a few essential features.
DANSS consists of 2500 polystyrene-based extruded scintillator strips ($1 \times 4 \times 100$~cm${}^3$) with a thin 
Gd-containing surface coating.  The strips are arranged in 100 layers of 25 strips. The strips in the adjacent layers are orthogonal. The detector is placed inside a composite shielding of copper, borated polyethylene, lead, and one more layer of borated polyethylene. It is surrounded by double layers of 3~cm thick scintillator plates to veto cosmic muons (the VETO system). 
The detector is installed under the core of a 3.1~GW$_{\rm th}$  industrial power reactor at the Kalinin Nuclear Power Plant (KNPP).
Light from the strip is collected with three wavelength-shifting Kuraray fibers Y-11 glued into grooves along the strip. 
The central fiber is read out with a Silicon PhotoMultiplier (SiPM).  The side fibers from 50 parallel strips
 are read out with a compact photomultiplier tube (PMT).

The high granularity of the detector allows reconstructing the energy of a positron from the IBD process without 
energy of the accompanying annihilation photons. Therefore the detector response does not depend on non-linearities
related to the detection of soft photons. The calibration is based on the comparison of the GEANT4 MC simulation with the detector response to energetic particles. The detector response to muons crossing the strips at different angles is linear with energy within 1\%. The experimental energy resolution for cosmic muon signals in the scintillator strips is somewhat worse than that from the MC calculation. Therefore, the MC estimates of the energy resolution are scaled up by the factor 17$\%/\sqrt{E}$ added in quadrature.

\begin{figure}[h]
\centering

\begin{minipage}[h]{0.3\linewidth}
\includegraphics[width=\linewidth]{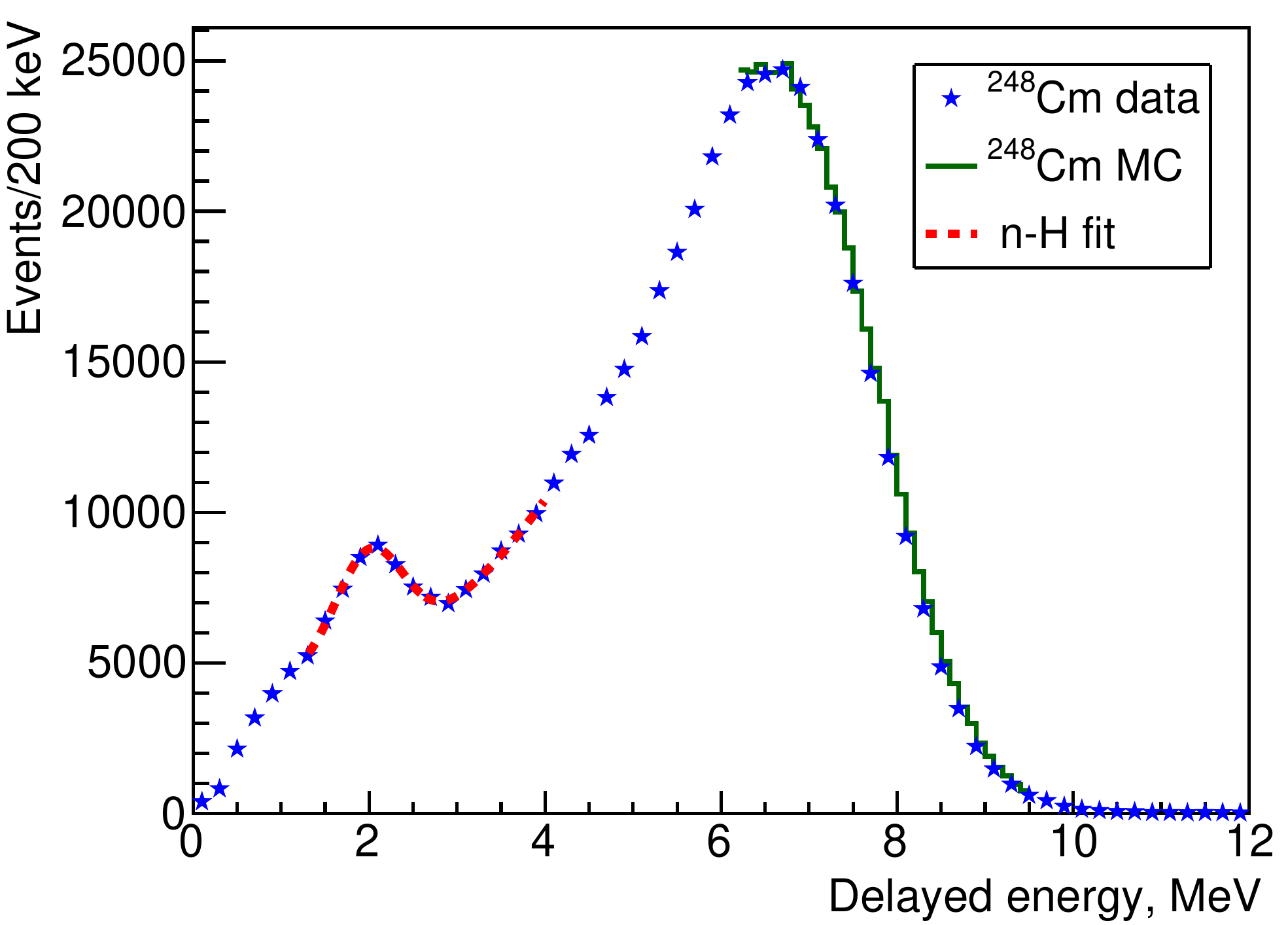}
\caption{\footnotesize Energy spectrum of the delayed signals measured with the $^{248}$Cm neutron source.}
\label{Cm}
\end{minipage} \hspace{2pc}%
\begin{minipage}[h]{0.6\linewidth}
\includegraphics[width=\linewidth]{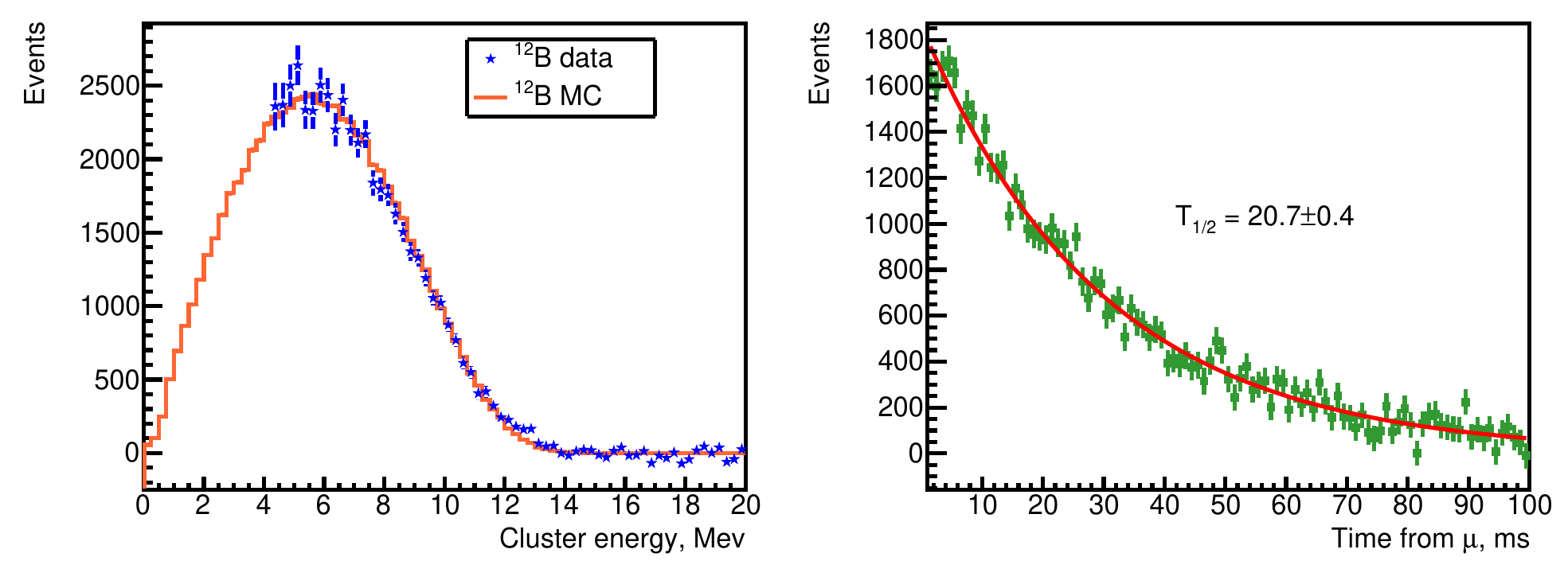}
\caption{\footnotesize Electron energy spectrum from $^{12}$B decays (left) and $^{12}$B decay time distribution(right). Red curves are MC predictions, while blue and green points are experimental data. }
\label{12B}
\end{minipage} 

\end{figure}

Fig.~\ref{Cm} shows the energy distribution of neutron capture signals from a $^{248}$Cm source placed at the center of the detector. 
Two peaks correspond to the neutron capture by Gd and by protons. 


Fig.~\ref{12B} shows the energy distribution for electrons from  $^{12}$B decays. $^{12}$B is produced by muons 
inside the detector. The spectrum is well described by the MC.


The fits of the high energy part of the Gd peak and $^{12}$B spectrum are used for anchoring the energy scale. It is fixed in such a way that the MC predictions for $^{12}$B and Gd decays are shifted from the data by -0.5\% and +1\% correspondingly. With such energy scale, MC predictions for $^{60}$Co and $^{22}$Na sources are shifted from the data by +2\% and +3\% correspondingly. 
 Most probably this non-linearity is related to a non-perfect description in MC of the low energy photon detection. In many other experiments such non-linearity was also observed and corrected for using empirical fits. In our case, such correction is not required since we do not use soft gammas for the positron energy determination. We measure the positron kinetic energy only.

Several improvements to the data processing and MC simulation have been made since our last 
publication \cite{DANSS-PLB}. PMT and SiPM signal waveforms are now used to determine $T_0$ and charge. The corresponding waveforms have been introduced into the MC simulation. The treatment of the Birks effect and Cherenkov radiation in MC was improved. Better corrections on the signal attenuation along the strips were introduced. They were determined directly from the data separately for the SiPM and PMT readout. The PMT signal coincidence with all 
SiPM signals is now required to suppress the SiPM noise. Annihilation photons are now required in case of a positron cluster consisting of one strip only. This requirement reduces the accidental background considerably as well as a fast neutron background. Two lowest scintillator strip layers have been added to the VETO system. The calibration frequency was increased to once in $\sim$ 25 minutes and once in 2 days for the SiPM gain calibration and the MIP calibration of all 2600 scintillator counters including counters of the VETO system correspondingly. 
Four times finer grid of points on the ($\Delta m_{14}^2$, $\sin^22\theta_{14}$) plane is now used.  A small bug in the oscillation pattern numbering was fixed. The last two changes resulted in a tiny shift in the best-fit point and the corresponding $\chi^2$ difference with the 3$\nu$ hypothesis for the same data sample as in the  published  results \cite{DANSS-PLB}. $\Delta m_{14}^2$ changed from 1.4~eV$^2$ to 1.33~eV$^2$, $\sin^22\theta_{14}$ changed from 0.05 to 0.056, and $\Delta\chi^2$ from 13.1 to 12.5.
Finally, the electron spectrum from $^{12}$B decays is now playing the key role in anchoring the energy scale.

These improvements resulted in the reduction of the accidental background from 71\% to 29\% and the cosmic muon background from 2.8\% to 1.9\% in both cases in comparison with the IBD rate at the top detector position. 
 The contribution of the accidental SiPM signals to the energy in the event was reduced from 100~keV to 5~keV.
 
\begin{figure}[h]
\begin{center}
\includegraphics[width=0.9\textwidth]{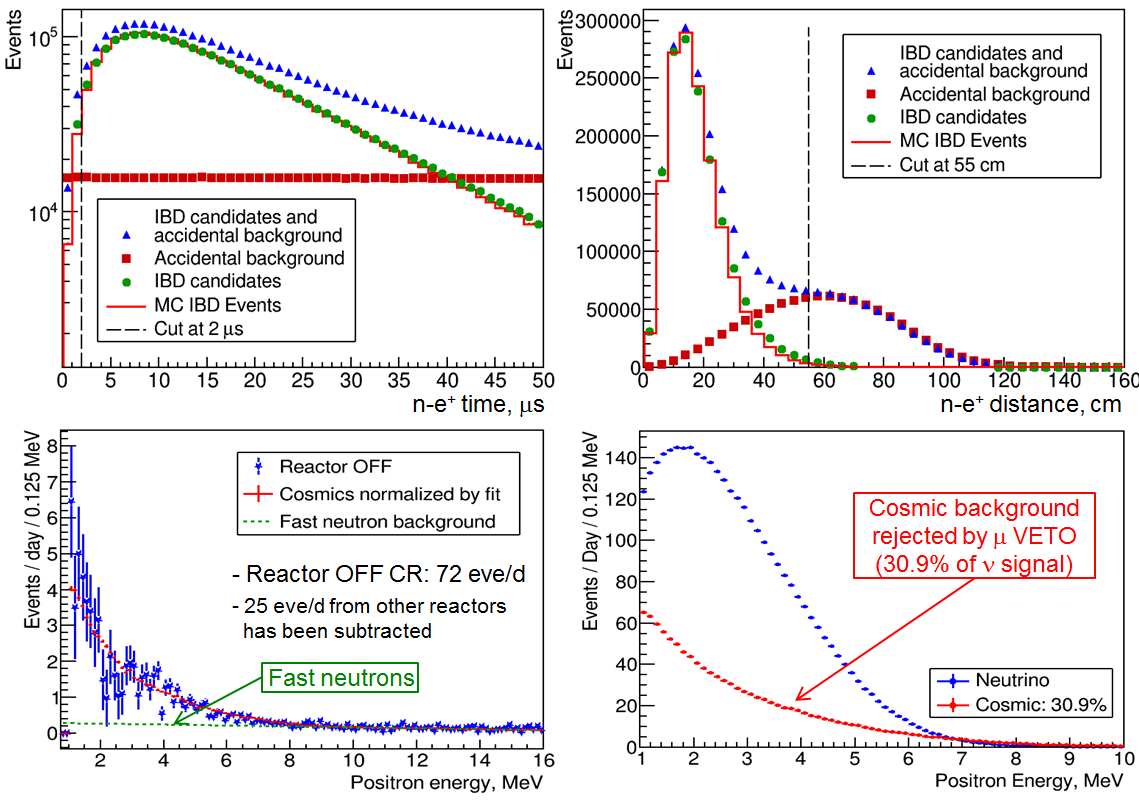}
\end{center}
\caption{\label{Background} Top: IBD candidate counting rate before (blue triangles) and after (green dots) accidental background (red squares) subtraction with the IBD signal MC predictions (red histogram) for $e^+$-n time and distance differences. Bottom left: Reactor OFF $e^+$ spectrum (blue points) fitted with cosmic muon background shape (red dots) and fast neutron background (green dashed curve). Bottom right: cosmic muon background rejected by the VETO system (red dots) in comparison with the IBD $e^+$ spectrum (blue dots).}
\end{figure}

 The background reduction and subtraction are illustrated in Fig~\ref{Background}. The accidental background is determined in a model-independent way directly from the data. We repeat the analysis for 16 time windows for the delayed signal well shifted from the prompt signal. The obtained accidental background is subtracted from all studied distributions (see e.g.~Fig.~\ref{Background}~(top)). Several very mild cuts have been applied to reduce the backgrounds \cite{DANSS-PLB}. Only the fiducial volume cut has a considerable inefficiency. The positron candidates are required to be at least 4 ~cm away from all detector edges. The background produced by cosmic muons has been mostly rejected by the VETO system. The inefficiency of the VETO system was determined using the reactor off data (see Fig.~\ref{Background}~(bottom)). The remaining background from cosmic muons is only 1.9\% of the IBD rate at the top position. The background from fast neutrons produced outside the detector shielding was estimated by linear extrapolation from the 10-16~MeV energy range to be 12~events/day only.  The background of 0.6\% from the nearby reactors was also subtracted. The cosmogenic $^9$Li and $^8$He background of about 4 events/day is simply ignored at the moment.

\begin{figure}[h]

\begin{minipage}[h]{0.3\linewidth}
\center{\includegraphics[width=0.95\linewidth]{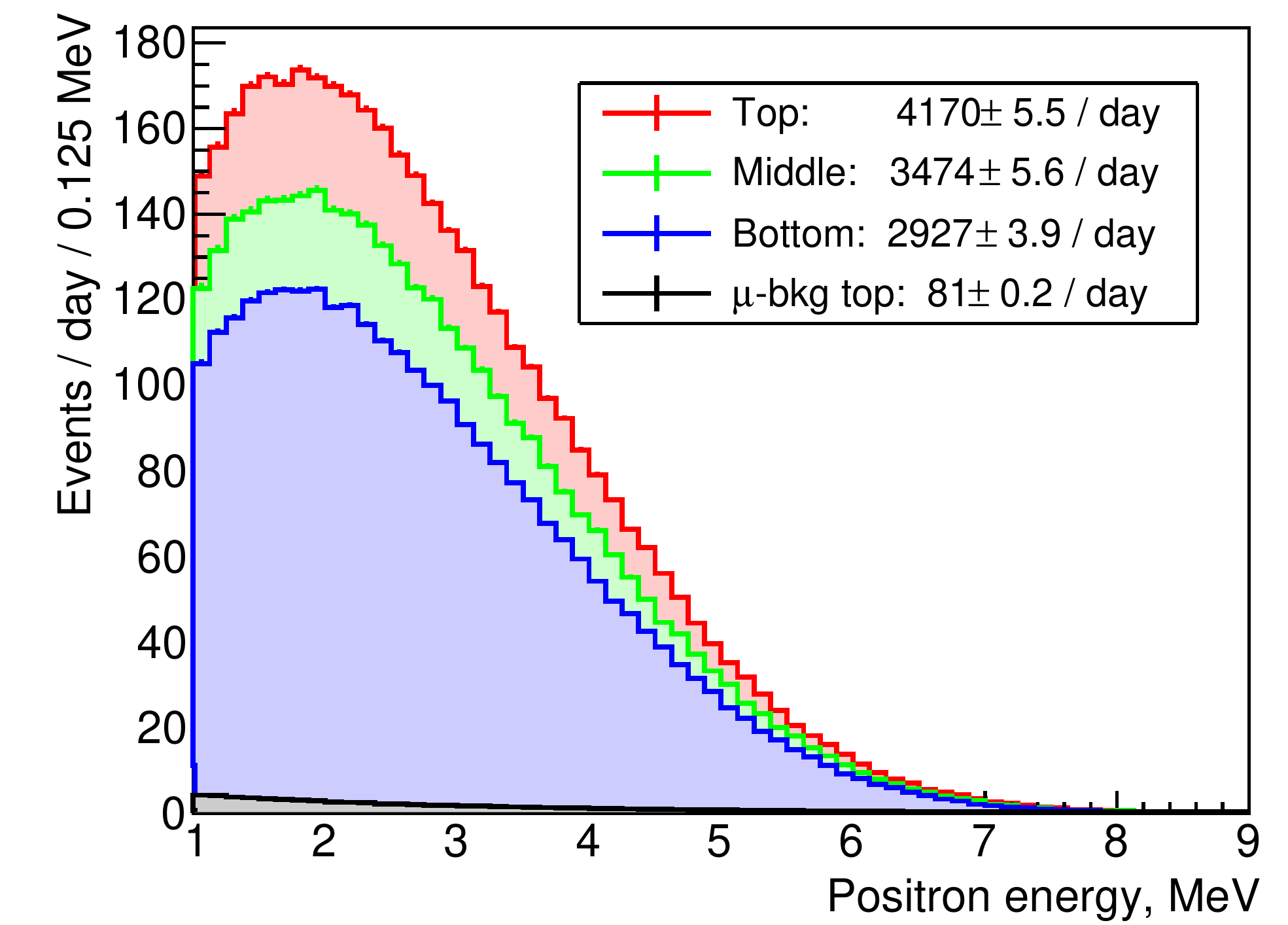}}
\end{minipage}
\hfill
\begin{minipage}[h]{0.3\linewidth}
\center{\includegraphics[width=0.95\linewidth]{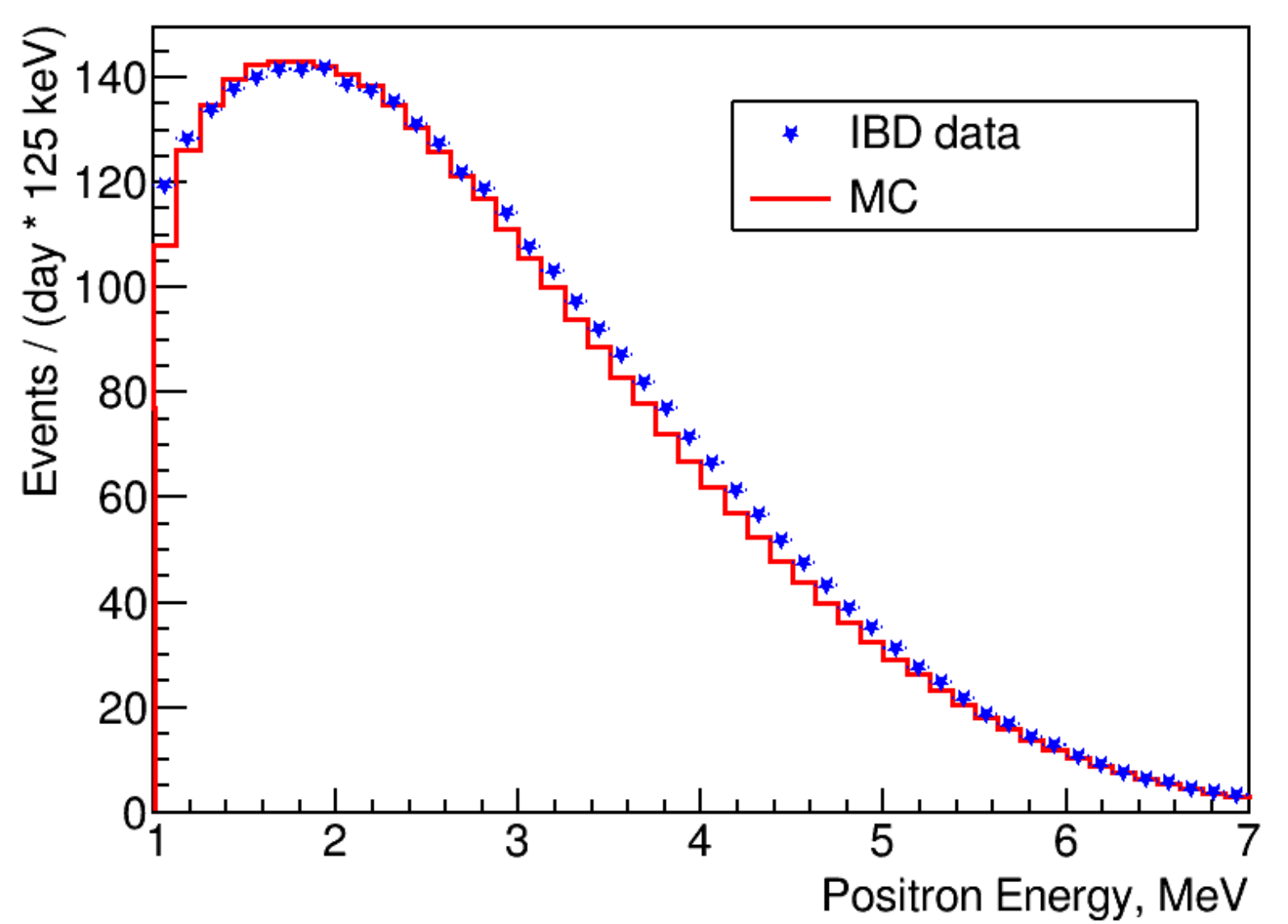} }
\end{minipage}
\hfill
\begin{minipage}[h]{0.3\linewidth}
\center{\includegraphics[width=0.95\linewidth]{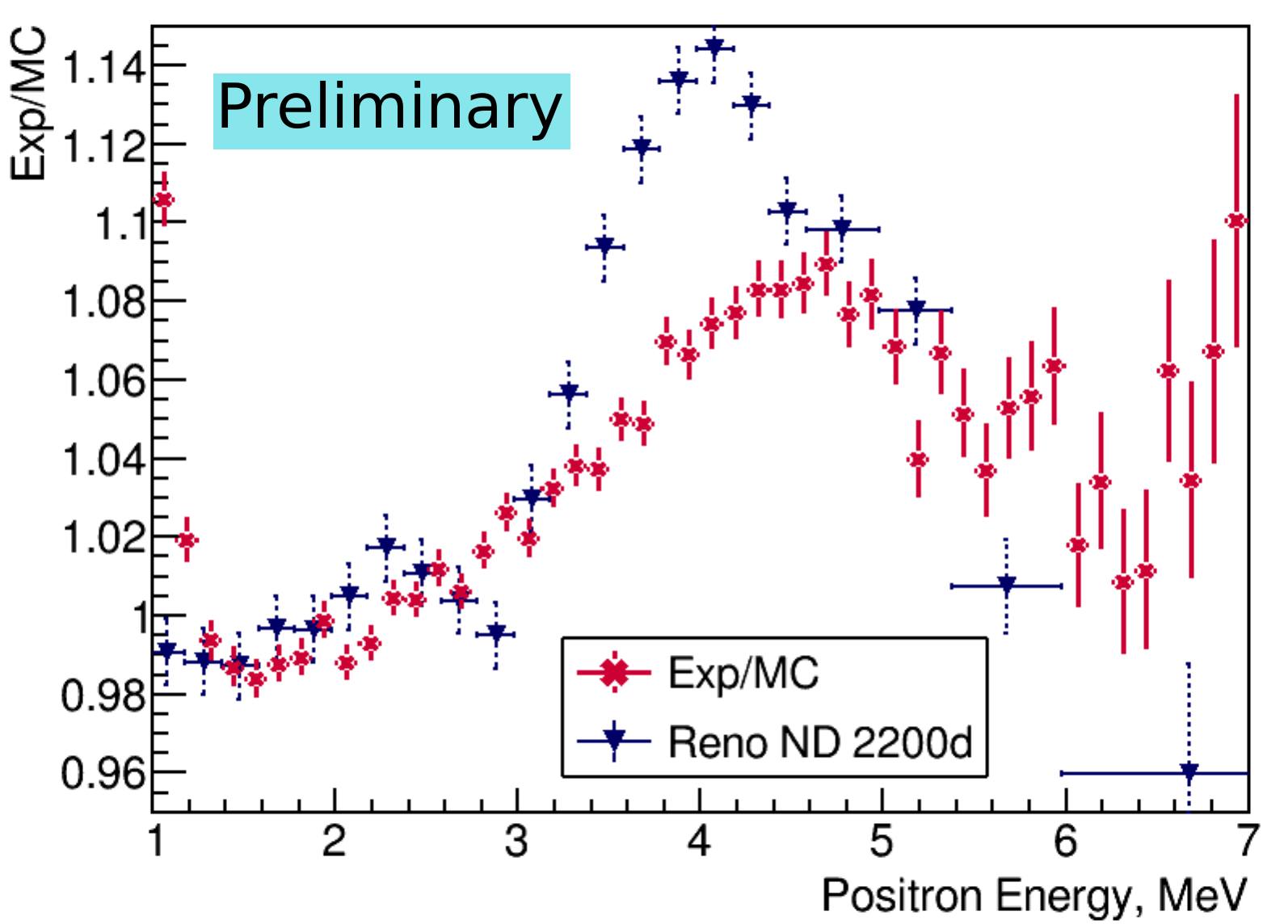} }
\end{minipage}
\caption{\label{Spectra} Left: IBD positron spectra at 3 detector positions and the cosmic muon background. Middle: Comparison of the IBD positron spectrum (blue stars) and the H-M model\cite{Mueller,Huber} (red histogram). Right: Ratio of the IBD positron spectrum and the H-M model(red dots) and the shifted RENO result\cite{RENO}.}
\end{figure}

 The obtained positron energy spectra for 3 detector positions (top, middle, bottom) are shown in Fig.~\ref{Spectra}~(left) with statistical errors only. 
The IBD counting rate exceeds four thousand events per day at the top detector position.
The positron energy does not include annihilation photons and hence it is 1.02~MeV lower than the usually used prompt energy.
The shape of the positron energy spectrum agrees roughly with the MC expectations based on the Huber-Mueller model (see Fig.~\ref{Spectra}~(center)) \cite{Mueller,Huber}. A detailed comparison shown in Fig.~\ref{Spectra}~(right) indicates the existence of some structure at the position where other experiments observe a bump. However, the shape of the spectrum depends strongly on the energy scale. Therefore we do not make any conclusion about the existence of the bump in our data until the completion of the studies of the systematic effects in the energy scale determination.

\begin{figure}[h]
\begin{minipage}[h]{0.48\linewidth}
\includegraphics[width=\linewidth]{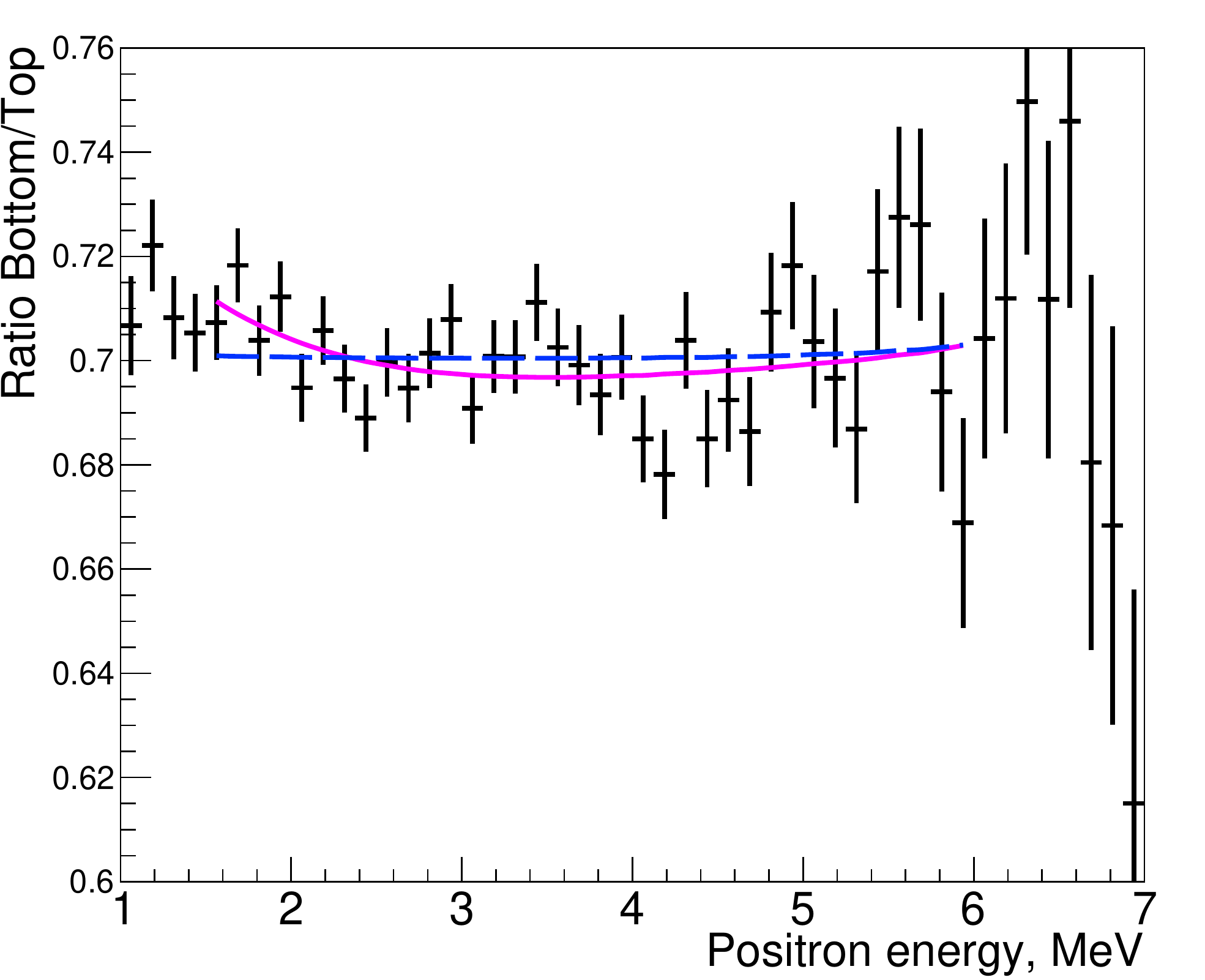}
\vskip 0.84cm
\caption{\label{Ratio}Ratio of positron energy spectra measured at the bottom and top detector positions ( statistical errors only). The dashed line is the prediction for 3$\nu$ case. The solid curve is the best fit in the $4\nu$ mixing scenario ($\Delta m_{14}^2$ = 0.35 eV$^2$, $\sin^22\theta_{14}$ = 0.15). 
}
\end{minipage}
\hfill
\begin{minipage}[h]{0.48\linewidth}
\includegraphics[width=\linewidth]{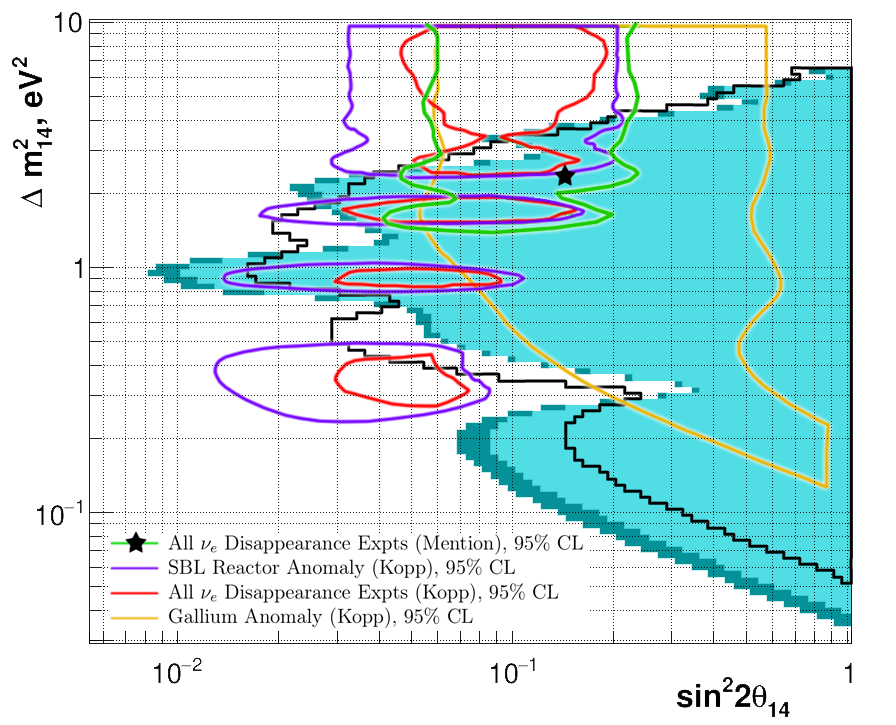}
\caption{\label{Exclusion}90$\%$(dark cyan) and 95\%(cyan) CL exclusion areas in the $\Delta m_{14}^2,~\sin^22\theta_{14}$ parameter space. The shaded areas represent our analysis. Curves show allowed regions from neutrino disappearance experiments \cite{Mention2011,contours}, and the star is the best point from the RAA and GA fit \cite{Mention2011}. The black curve is the 90 \%CL sensitivity of the DANSS experiment.}
\end{minipage}
\end{figure}

 The difference in $\chi^2$ between the 4$\nu$ and 3$\nu$ hypotheses was 13.1 in the old data set \cite{DANSS-PLB}. 
This created some excitement although we clearly stated in the paper that this difference does not take into account systematic effects and postponed the study of its statistical significance until the collection of more data. The new data set that is 1.5 times larger does not show any significant sign of 
$\anti_particle\nu$ oscillations. The largest $\Delta\chi^2$ between the 4$\nu$ and 3$\nu$ hypotheses is only 2.3 if one takes into account the systematic uncertainties that include the reduction of the fit range to 1.5-6~MeV (it is only 4.6 in the 1-7~MeV range and without the systematic uncertainties). The same reduction of the fit range was used in our published analysis \cite{DANSS-PLB} for the estimation of the systematic
uncertainties that reduce the exclusion range on the $\Delta m_{14}^2$, $\sin^22\theta_{14}$ plane.   
Other studied systematic effects include variations in the energy scale ($\pm 2\%$) and energy resolution ($\pm10\%$) as well as variations of the level of the cosmic muon background ($\pm25\%$) and a flat background ($\pm30\%$ of the fast neutron background). The full data set also does not show a statistically significant sign of $\anti_particle\nu$ oscillations (see Fig.~\ref{Ratio}). The largest $\Delta\chi^2$ between the 4$\nu$ and 3$\nu$ hypotheses is   
7.8 that corresponds to the 1.8$\sigma$ confidence level (CL) only. 

The exclusion area in the sterile neutrino parameter space was calculated using the Gaussian \cls method \cite{CLS} assuming only one type of sterile neutrinos. For a grid of points in the $\Delta m_{14}^2,~\sin^22\theta_{14}$ plane predictions for the ratio $R^{pre}(E)$  of positron spectra at the bottom and top detector positions were calculated. Calculations included the MC integration over the $\anti_particle\nu_e$  production point in the reactor core,  $\anti_particle\nu_e$  detection point in the detector, and positron energy resolution. The $\anti_particle\nu_e$  production point distributions in the reactor core were provided by the KNPP for different periods. The distribution averaged over the campaign was used in the calculations. It was checked that this approximation practically did not influence the final results. Predicted positron spectra at different detector positions were normalised using the observed numbers of IBD events. 

The obtained theoretical prediction for a given point on the $\Delta m_{14}^2,~\sin^22\theta_{14}$ plane was compared with the prediction for the three neutrino case using the Gaussian \cls method for the exclusion area estimation. The difference in $\chi^2$ for the two hypotheses $\Delta\chi^2 = \chi^2_{4\nu} - \chi^2_{3\nu}$ was used for the comparison.  
The $\chi^2$ for each hypothesis was constructed using 36 data points $R^{obs}_i$ in the 1.5-6~MeV positron energy range and minimized over nuisance parameters (systematic effects) using all combinations of their discrete values mentioned above:

\begin{equation}
\label{eq2}
\chi^2 = \sum_{i=1}^N(R^{obs}_i-R_i^{pre})^2/\sigma_i^2,
\end{equation}
where $R^{obs}_i$ ($R^{pre}_i$) is the observed (predicted) ratio of $\anti_particle\nu_e$ counting rates at the two detector positions and $\sigma_i$ is the statistical standard deviation of $R^{obs}_i$. The $\chi^2$ does not depend on the integral IBD event rate dependence on the distance from the reactor core because of $R^{pre}$ normalisation. Only differences in the positron energy spectrum shapes are considered. 

 The oscillations due to the known neutrinos were neglected since at such short distances they do not change the $\anti_particle\nu_e$ spectrum. 
 The procedure was repeated for all points of the grid in order to get the whole exclusion area that expands 
 further the area excluded in our previous analysis \cite{DANSS-PLB}. The sensitivity of the experiment was improved by a factor of $\sim1.4$.  
Fig.~\ref{Exclusion} shows the obtained excluded area in the $\Delta m_{14}^2,~\sin^22\theta_{14}$ plane together with some initial expectations based on GA and RAA. Our results exclude a large and very interesting part of the sterile neutrino parameter space. Unfortunately, the sensitivity of the present analysis is low at large $\Delta m_{14}^2$. Therefore we can not check the NEUTRINO-4 claim \cite{Neutrino-4}.

\section*{Acknowledgments}
We appreciate the permanent assistance of the KNPP administration and Radiation and Nuclear Safety Departments. The detector construction was supported by the Russian State Corporation ROSATOM (state contracts H.4x.44.90.13.1119 and H.4x.44.9B.16.1006). The operation and data analysis became possible due to the valuable support from the Russian Science Foundation grant 17-12-01145. The preparation of this paper was supported by the Russian Federal Government grant 14.W0331.0026.


\end{document}